\documentstyle{article}
\everymath{\rm }
\everydisplay{\rm }

\begin{document}
\begin{center}
{\LARGE Open-Flux Solutions to the \\ [.125in] Quantum
Constraints
for\\ [.125in] Plane
Gravity Waves} \\ [.25in]
\large Donald E. Neville \footnote{\large e-mail address:
nev@vm.temple.edu }\\Department of Physics \\Temple University
\\Philadelphia 19122, Pa. \\ [.25in]
November 6, 1995 \\ [.5in]
\end{center}
\newcommand{\E}[2]{\mbox{$\tilde{{\rm E}} ^{#1}_{#2}$}}
\newcommand{\A}[2]{\mbox{${\rm A}^{#1}_{#2}$}}
\newcommand{\Np}{\mbox{${\rm N}'$}}
\newcommand{\Etwo}{\mbox{$^{(2)}\!\tilde{\rm E} $}}
\newcommand{\Etld }{\mbox{$\tilde{\rm E}  $}}
\def \ut#1{\rlap{\lower1ex\hbox{$\sim$}}#1{}}
\newcommand{\phst}{\mbox{$\phi\!*$}}
\newcommand{\bea}{\begin{eqnarray}}
\newcommand{\eea}{\end{eqnarray}}
\newcommand{\be}{\begin{equation}}
\newcommand{\ee}{\end{equation}}
\newcommand{\rta}{\mbox{$\rightarrow$}}
\newcommand{\EQ}[1]{equation~(\ref{eq:#1})}
\Roman{section}
\large
\begin{center}
{\bf Abstract}
\end{center}
The metric for plane gravitational waves is quantized within the
Hamiltonian framework, using a
Dirac constraint quantization and the self-dual field variables
proposed by Ashtekar.  The z axis
(direction of travel of the waves) is taken to be the entire real
line rather than the torus (manifold
coordinatized by (z,t) is RxR rather than $S_1$ x R).
Solutions to the constraints are
proposed; they involve open-ended flux lines running along the
entire z axis, rather
than closed loops of flux.  These solutions are annihilated by
the
constraints at all interior points
of  the z axis.  At  the two boundary points, the Gauss
constraint
does not annihilate the
solutions, because of  the presence of open-ended flux lines at
the
boundaries.  This result is in
sharp contrast to the situation in the general, 3+1 dimensional
case without planar symmetry,
where the Gauss constraints do not contribute at the boundaries
because the Lagrange multipliers
for the Gauss constraints fall to zero at spatial infinity.  In
the
planar symmetry case, the
Lagrange multiplier for rotations about the internal Z axis
survives at the boundary.   The
constraints are found to annihilate the solutions when classical
matter terms are added to the
Hamiltonian (so that flux lines are terminated on the matter).
The
holonomy matrices used in
the solutions are generated by the (2j+1) dimensional
representation of SU(2), where j may be
any spin, not necessarily j = 1/2.  In this respect the solutions
resemble the symmetric states, or spin network
states recently constructed by Rovelli and Smolin in loop space.
The Rovelli-Smolin area
operator for areas in the xy plane is constructed and applied to
the solutions, with puzzling
results. \\[.125in]
PACS categories: 04.60, 04.30

\section{Introduction}

     The connection-triad variables introduced by Ashtekar
\cite{Ash87} have
simplified the constraint equations of quantum gravity; further,
these variables suggest that in the future we may be able to
reformulate gravity in terms of non-local holonomies rather than
local field operators \cite{RovSmo,GambTri}.  However, the new
variables are
unfamiliar, and it is not always clear what they mean physically
and geometrically.  In particular, it is not clear what operators
or structures correspond to gravity waves.  Although the quantum
constraint equations are much simpler
in the new variables, and solutions to these equations have been
found \cite{RovSmo,knotsol}, it is not clear whether any of
these solutions contain gravitational radiation.

     This is the second of a series of three papers which search
for operator signatures for gravitational radiation
by applying the Ashtekar
formalism to the problem of plane gravitational waves.  Paper I
in the series \cite{I} constructed classical constants of the
motion for
the plane wave case, using the more familiar geometrodynamics
rather than Ashtekar connection dynamics.  Paper
II, the present paper, switches to connection dynamics, carries
out a quantization of the plane wave metric,
proposes solutions to the quantum constraints, and
proves that the constraints annihilate these solutions except at
boundary points.  Paper III, which is in preparation, will
propose operator signatures for gravitatonal radiation and apply
those operators to the solutions constructed in the present
paper.

     The phrase "plane gravitational wave" is used in this paper
as a shorthand description for a
specific class of metrics.  "Plane"
means the metrics possess two commuting spacelike Killing
vectors.  I choose x and y coordinates so that these
vectors are unit vectors pointing in the x and y directions.
\begin{equation}
     k^{(x)} = \partial _x; k^{(y)} = \partial _y.
\label{1.0}
\end{equation}
The words "gravitational wave"
may be defined precisely in a number of ways \cite{EhlK} .
Szekeres \cite{Sz} suggests a definition which has considerable
intuitive
appeal: the system possesses an orthonormal tetrad consisting
of $k^{(x)}\ and\ k^{(y)}$ plus two null vectors which are
hypersurface orthogonal.
The two null vectors may be interpreted as ray vectors for
left-and right-moving radiation along the z axis.   The two
hypersurfaces are the corresponding wave fronts.

     It is possible to choose the
coordinates (z,t) so that these wave fronts have simple
equations $ct \pm z = constant$, but I follow the usual
philosophy in the quantization literature and leave (z,t)
arbitrary.  When the  (x,y) coordinates are gauge-fixed,
the system becomes simpler than the full four-dimensional
case.  If  the (z,t) coordinates are gauge-fixed as well, then
the system
becomes too simple and does not illuminate the full case.

      I quantize the theory in section 2 and derive wave
functional solutions to
the constraints in section 3.  These
solutions resemble Rovelli-Smolin $T^n$ operators,
 strings of triad operators separated by holonomies
\cite{RovSmo,JRS}, except that
the solutions fill the entire z axis (they are open, rather than
closed flux loops) and each triad is integrated over z so as to
guarantee invariance under z-diffeomorphisms.

     Also, the solutions utilize SU(2) generators which are
(2j+1)x(2j+1)
dimensional, rather than the usual 2x2 Pauli matrices.  A 2x2
generator can be viewed as acting upon a single flux line, while
a (2j+1)x(2j+1) generator can be viewed as acting upon 2j flux
lines, which have been totally symmetrized to give a total spin
state of spin j.  These new solutions may be closely related
to the symmetric states, or "spin network" states
recently constructed by Rovelli and Smolin in loop space
\cite{symmsta}.  I
consider the introduction of the (2j+1)x(2j+1) generators as the
most important technical innovation of the present paper.

     In quantizing the theory, I follow earlier work by Husain
and Smolin \cite{HSm} with one signifigant exception: I
incorporate a suggestion due to Teitelboim \cite{Teinorm} and
rescale the scalar constraint.  In order to understand the
advantages of this rescaling, consider the standard Dirac
requirement
that a wave functional in the physical subspace must be
annihilated by an arbitrary linear combination of the
constraints:
\begin{eqnarray}
     0 &=& [\ut{N} H + N^z H_z + N_G H_G]\psi \nonumber \\
       &=& [(\ut{N} \E{z}{Z})(H /\E{z}{Z}) + N^z H_z
               + N_G H_G]\psi \nonumber \\
       &\equiv& [(\Np)(H_S) + N^z H_z + N_G H_G]\psi .
\label{eq:1.3}
\end{eqnarray}
H and $H_z$ are the usual Asktekar scalar constraint and
generator of z diffeomorphisms; $H_G $ is the Gauss constraint,
the  generator of rotations around the internal Z axis.  The
planar symmetry has been used to gauge away the generators
of x and y diffeomorphisms and generators of internal X and Y
rotations.  \E{z}{Z} is the usual Ashtekar momentum variable, the
densitized inverse triad $e e^z_Z$.  The seemingly
innocuous rescaling from H to $H_S$ in \EQ{1.3} has profound
consequences for
the algebra of constraints.  Consider in particular the
scalar-scalar commutator \cite{Alect}
\begin{eqnarray}
     [H(x),H(x')] &=& [H_i(x) gg^{ij}(x) + H_i(x') gg^{ij}(x')]
\partial _j \delta (x-x') +\cdots \nonumber
\\
       &\rta & [H_z(z) (\E{z}{Z}(z))^2 +
          H_z(z') (\E{z}{Z}(z'))^2] \partial _z \delta (z-z') +
\cdots.
\label{eq:1.4}
\end{eqnarray}
The $\cdots$ indicates  terms proportional to the Gauss
constraints, omitted for simplicity
because they do not affect the argument.
The first line of   \EQ{1.4} gives the commutator in the general,
3+1
dimensional case; the second line specializes to
planar symmetry.  For a consistent quantization, the
right-hand side of each line must annihilate $\psi $, since the
left-hand side does.  In the general case, for most choices of
operator ordering of the constraints $H$ and $H_i$, the $gg^{ij}$
factor occurs to the right of the $H_i$ and prevents the latter
from annihilating $\psi$.
(It is possible to find
orderings for which the $gg^{ij}$ factor occurs to the left, but
these orderings
do not permit $H_i $ to be interpreted as a generator of
diffeomorphisms in the ith direction \cite{OptrOrd}.)   Now
rescale as at \EQ{1.3}.
The rescaling  modifies the commutator.
\[
     [H_S(z),H_S(z')] = [H_z(z) +
                H_z(z')] \partial _z \delta (z-z')+\cdots.
\label{eq:1.5}
\]
The unwanted $gg^{ij}\ or\  \E{z}{Z}$ factors have disappeared.
As a result, in section 2, I will be able to pick a factor
ordering
which makes the constraint algebra  consistent and simultaneously
permits $H_z$ to be interpreted as a generator of
diffeomorphisms.

     This is non-trivial progress; but it is not yet enough to
guarantee a consistent quantization.  The right-hand side of
\EQ{1.5} could contain Schwinger terms, c-number terms which
spoil consistency because
they  do not annihilate $\psi$.  If one had a
complete set of solutions, presumably one could check for the
presence of Schwinger terms by computing every matrix element of
the commutator on the left in \EQ{1.5}.  Unfortunately, no such
complete set is available.  However, Kucha\v{r} \cite{Kscal} has
demonstrated that Schwinger terms are absent in a model which is
very close to the present case, yet still  solvable.  The model
is a free massless scalar field on a flat 1+1 dimensional
manifold.  This model is made generally covariant by adding
"embedding variables", plus scalar and vector constraints which
guarantee that the new variables are not dynamical.  Kucha\v{r}
rescales the constraints so that the algebra of constraints
gives \EQ{1.5} for the scalar-scalar commutator.  Since he has a
complete
mode expansion, he is
able to calculate Schwinger terms, and he finds no such terms in
the constraint algebra.  (The Lie algebra of the conformal group
acquires Schwinger terms, in a manner familiar from string
theory; but this conformal algebra is not the  same as the Dirac
algebra of constraints.  Only Schwinger terms in the latter
algebra destroy consistency.)

     Since $H_S = H/\E{z}{Z} $, and H is polynomial, the new
scalar constraint $H_S $ is rational.  I view this complication
as a price which must be paid; but it is a small price,
considering what one gets in return.   The constraints close,
which they should as a matter of principle; and the quantum
theory possesses
the diffeomorphism invariance  characteristic of the classical
theory \cite{HKTei}.

     Nevertheless, at some point a price must be paid: how
does one define the inverse operator $ (\E{z}{Z})^{-1)} $?  All
but one of the terms in the Ashtekar scalar constraint H contain
a factor of \E{z}{Z}; therefore
classically \E{z}{Z} completely cancels out of all but one of the
terms in $H_S = H/\E{z}{Z} $.  I will assume this cancellation
occurs also in the quantum theory, so that $H_S $ contains only
one term with an \E{z}{Z} in the denominator.  The numerator of
this particular term vanishes, when acting on the solutions
constructed in this paper.  Therefore, so long as the action of $
(\E{z}{Z})^{-1} $ on $\psi $ does not give an infinity,
$(\E{z}{Z})^{-1} $ is not a problem in
the present paper, and one can
postpone a complete investigation of the inversion problem.  In
Appendix C,  I determine enough of the action of $ (\E{z}{Z})^{-
1} $ to show that this operator does not give an infinity, at
least when acting on the solutions constructed in this paper.

     In three spatial dimensions it is usual to place the
boundary surface at spatial infinity.  Bringing the surface at
infinity in to finite points is a major change, because at
infinity the metric goes over to flat space, and flat space is a
considerable simplification.  In the present case (effectively
one
dimensional because of the planar symmetry)  the space does {\it
not} become flat at z goes to infinity, and nothing is lost by
considering an arbitrary location for the boundary surface.  The
"surface" in one dimension is of course just two points (the two
endpoints of a
segment of the z axis).  These points will be taken to
be a finite distance from the origin.  The result that the space
does not become flat as z goes to infinity will be established in
section 4; but for now, note that this result agrees with one's
intuition from Newtonian gravity, where the potential in one
spatial dimension due to a bounded source does
not fall off, but grows as z at large z.

     Since the solutions are constructed from open flux tubes,
there is internal "Gauss charge" exiting through the surface
points.
In 3+1 dimensions, this exiting flux has no effect on
dynamics, since the three Gauss charge operators $H_I$ in the
Hamiltonian are multiplied by  Lagrange multipliers which fall to
zero at spatial infinity.  The situation is different in the
planar case.
The high symmetry allows two of the
Gauss constraints
to be solved and eliminated from the theory.  From the
discussion of the behavior at spatial infinity, given in sections
4 and 5, the Lagrange multiplier $N_G $ for the surviving Gauss
constraint $H_Z \equiv H_G $ is {\it finite } at infinity.
This means that the quantum
Hamiltonian annihilates the solutions at finite z, but not at
the boundary, because (the
multiplier $N_G $ is finite there and) the $H_G$ term in the
Hamiltonian is non-zero when acting on the exiting flux.

     At this point one can think of (at least) four ways to
proceed.  (1)
Ignore the difficulties at infinity and study the waves at finite
z only.  (2) Insert classical matter sources proportional to
$\delta(z-z_b),\ z_b$ the boundary points,  into the Gauss
constraint $H_G$, so that
the modified constraint annihilates the solutions.  The exiting
flux lines can be visualized as terminating on this matter.  (3)
Use second quantized
matter sources : introduce Fermion fields into the fundamental
Lagrangian and terminate the flux tubes on
Fermion fields
\cite{MTR}.  (4)  Search for closed flux line solutions.

     The solutions (3)-(4) are the most satisfying in principle.
Even in practice, presumably enough is known about Fermions to
implement procedure (3) using the ansatz of section 3 for the
gravitational part of the solution, although the calculation
would be technically intricate.  Procedure (4) probably requires
a radically new ansatz; see the comments at the end of section 3.

     In this paper and the succeeding one I shall use
procedures (1) or  (2).   These two procedures
are identical in spirit: both ignore the difficulties at
infinity, and study the waves at
finite z.   Section 5 of the present paper constructs the matter
sources needed for
procedure (2).

     Procedures (1)-(2) have the virtue of
simplicity, and indeed (1) is essentially the procedure employed
in classical electromagnetism, where normally one studies a
monochromatic plane wave.  In electromagnetism there are no
problems with a
Gauss charge, but the wave stretches from $z = + \infty$ to $z =
-\infty$
and has infinite energy.  The unphysical nature of  this wave at
infinity
does not stop one from studying the
wave at finite z and learning a great deal.

     In the electromagnetic case there is an analog of
procedures (2)- (3), the inclusion of matter.  If one
wishes, one can eliminate the infinite energy by  confining the
radiation between two parallel metal
plates.  In practice, this procedure is never invoked until one
wishes to discuss more advanced
topics such as waveguides or Fresnel relations at conducting
boundaries.  The introduction of metal boundaries at any earlier
point
would be a distraction.  I take the
same point of view here: it is important to introduce realistic
matter at
some point; but initially the focus should remain on the
radiation.

     The rest of this paper is organized as follows.  Section 2
quantizes and regulates the theory.  Section 3 proposes an ansatz
for the
solutions, and proves that the Hamiltonian annihilates the
solutions at all finite points away from boundaries.  Behavior at
boundaries is largely ignored in section 3, and in particular the
surface terms in the Hamiltonian are ignored.  Section 4
constructs these surface terms.  Section 5 returns to the
solutions of section 3 and shows that the Hamiltonian (now with
surface terms included) does not annihilate the solutions at
boundaries. Section 5 also implements procedure (2): matter
sources are added at the
boundaries so that the Hamiltonian does annihilate the solutions
everywhere. Section 6 discusses generalizations
and directions for further research.

     Appendix A explains the
sign conventions I have adopted in
reducing the Ashtekar formalism from its covariant,
four-dimensional form to 3+1
dimensions.  The
convention normally adopted in the literature \cite{Rovlect} has
unintended consequences which I believe many physicists would
prefer to avoid.

     My notation is typical of papers based upon the Hamiltonian
approach with concomitant 3+1 splitup (except for the sign
conventions
explained in Appendix A).  Upper case indices A,
B, $\ldots $,I, J, K, $\ldots$ denote local Lorentz indices
("internal" SU(2) indices) ranging over X, Y, Z only.  Lower case
indices a, b, $\ldots $, i, j, $\ldots $ are also three-
dimensional and denote global coordinates on the three-manifold.
Occasionally the formula will contain a field with a superscript
(4), in which case the local Lorentz indices range over X, Y, Z,
T and the global indices are similarly four-dimensional; or a
(2), in which case the local indices range over X, Y (and global
indices over x, y) only.  The (2) and (4) are also used in
conjunction with determinants; e.\ g., g is the usual 3x3
spatial determinant, while\ $^{(2)}e$ denotes the determinant of
the 2x2 X, Y subblock of the triad matrix $e^A_a$. I use Levi-
Civita symbols of various dimensions: $\epsilon _{TXYZ} =
\epsilon _{XYZ} = \epsilon _{XY} = +1$.  The Minkowski metric
convention is
$\eta_{TT} = -1$.  The basic variables of
the Ashtekar approach are an inverse densitized triad \E{a}{A}
and a complex SU(2) connection \A{A}{a}.
\begin {eqnarray}
     \E{a}{A}& =& \sqrt g e^a_A; \\
\label{eq:1.1}
     [\E{a}{A},\A{B}{b}]&=& -\hbar \delta (x-x') \delta ^B_A
                         \delta ^a_b.
\label{eq:1.2}
\end{eqnarray}
The local Lorentz indices are vector rather than spinor.
Strictly
speaking the internal symmetry is O(3) rather than SU(2), gauge-
fixed to O(2) rather than U(1).

\section{Quantization}

     This section begins with a brief discussion of  the
simplifications which ensue
when the (x,y) coordinates are fixed, and follows this with a
description of the Hamiltonian for
plane waves.   Then the system is quantized in a standard way, by
replacing certain fields with functional derivatives.  Issues
such
as factor ordering, regularization, and closure of constraints
are discussed.

     The high degree of symmetry associated with the two Killing
vectors allow Husain and Smolin to solve and eliminate four
constraints (the
x and y vector constraint and the X and Y Gauss constraint) and
correspondingly eliminate four pairs of  (\E{a}{A}, \A{A}{a})
components \cite{HSm}.  The 3x3 \E{a}{A}\
matrix then assumes a block diagonal form, with one 1x1 subblock
occupied by \E{z}{Z}\, plus one 2x2 subblock which contains all
\E{a}{A} with a = x,y and A = X,Y.  The 3x3 matrix of connections
\A{A}{a}\ assumes a similar block diagonal form.  None of the
surviving fields depends on x or y.

     After these simplifications, the total Hamiltonian reduces
to a linear combination of the three surviving constraints,
\begin{eqnarray}
     H_T &=& \Np [\epsilon _{MN}\E{x}{M} \E{y}{N}
          (\E{z}{Z})^{-1} \epsilon _{AB} \A{A}{x} \A{B}{y}
     + \epsilon _{MN} \E{b}{M} {\rm F}^N_{zb}]  \nonumber \\
     & & + iN ^z \E{b}{M} {\rm F}^M_{zb} \nonumber \\
          & & -i{\rm N_G} [\partial _z\E{z}{Z} - \epsilon_{IJ}
                     \E{a}{I} \A{J}{a} ] + S.T.\nonumber \\
     &\equiv & \Np H_S + N^zH_z + {\rm N_G}H_G +
S.T.,\label{eq:2.1a}
\end{eqnarray}
where\begin{equation}
     {\rm F}^N_{zb} = \partial _z\A{N}{b}
                          -\epsilon _{NQ}\A{Z}{z} \A{Q}{b}.
\label{eq:2.1b}
\end{equation}
$H_S, H_z$, and $H_G$ are the surviving scalar, vector, and Gauss
constraints.  Strictly speaking these are Hamiltonian densities;
for simplicity I have suppressed
an integration over the z axis.  S.T. denotes surface terms
evaluated at the two endpoints on the z axis.  The detailed form
of these terms will not be needed until section 4.  In
equation~(\ref{eq:2.1a}) the Lagrange multiplier \Np\ and the
scalar constraint $H_S $ are rescaled versions of the usual
Lagrange multiplier \ut{N}  and Ashtekar scalar constraint H, as
at
\EQ{1.3}.

     In the general, three-dimensional case, often one chooses a
wave functional depending on the A's, $\psi = \psi [A]$.  Since
the A's are connections, one is led to consider holonomies, and
then a loop space representation \cite{RovSmo,GambTri}.  In the
planar case, only
\A{Z}{z} is a connection, and there is no special incentive to
use any A except \A{Z}{z} in the wave functional.  I shall
choose a wave functional depending on this connection plus the
fields
\begin{equation}
     \E{a}{\pm} = [\E{a}{X} \pm i\E{a}{Y} ]/\sqrt{2} ,
\label{eq:2.1d}
\end{equation}
where a = x,y.  These fields are eigenstates of the surviving
gauge invariance O(2) or U(1) generated by $H_G$.  Alternatively,
one could use
\A{\pm }{a} eigenstates in the wave functional, but the weight
one fields in equation~(\ref{eq:2.1d}) will be signifigantly more
convenient because they are densities and can be made
diffeomorphism invariant simply by integrating them over z.

     The one connection field
in the wave functional, \A{Z}{z}, will be incorporated
into holonomy matrices in the usual way,
\begin{equation}
     M(z_2,z_1) = \exp [i\int_1^2 \A{Z}{z} {\rm S}_Z dz],
\label{eq:2.2}
\end{equation}
M may be visualized as a flux line extending along the z axis
from $z_1\ to\ z_2$. There is an explicit factor of i because
$S_Z
$ is the usual (Hermitean) generator.

     Now The wave functional $\psi$ depends on the commuting set
[\E{a}{\pm},\A{Z}{z}] which I will call the Q set, and does not
depend on the
conjugate P set [\A{\mp}{a}, \E{Z}{z}].  One may obtain a
representation
of the commutators~(\ref{eq:1.2}) in the usual way, by writing
the P's as functional derivatives.
\begin{eqnarray}
     \A{\pm}{a} &=& \hbar \delta/\delta \E{a}{\mp}; \nonumber \\
     \E{Z}{z}   &=& -\hbar \delta/\delta \A{Z}{z}.
\label{eq:2.3a}
\end{eqnarray}
(If the pattern of $\pm$ signs seems strange, note that the two
dimensional Kronecker delta in \EQ{1.2} has only off-diagonal
elements when
expressed in terms of O(2) eigenstates: $\delta_{\pm \mp} = +1$.)

     Tentatively I choose to factor-order  the Hamiltonian with
P's to the
right, Q's to the left, so that the $H_z $ constraint
generates diffeomorphisms along z.  (More precisely, the $H_z$
constraint generates diffeomorphisms after a piece proportional
to $\A{Z}{z} H_G $ is added \cite{Alect}.)  This ordering has
been adopted already in
equation~(\ref{eq:2.1a}).

     The Hamiltonian must now be regulated, since it contains
products
of functional derivatives evaluated at the same point.  These
products can act on the wave functional and produce undefined
products of delta functions evaluated at the same point.
There are no
products of functional derivatives in the Gauss and
diffeomorphism constraints, so that only the scalar constraint is
in need of regularization.  Every term in the original Ashtekar
scalar
constraint H contains a product of two functional derivatives,
but  the
division by \E{z}{Z}\ cancels one functional derivative from
most terms, leaving behind a single functional derivative
which no longer requires regularization.  In fact the only term
in $H_T $ needing regularization is the $(\E{z}{Z})^{-1} $ term
in $H_S $.  In the general, 3+1 dimensional case it is very
difficult to find a regularization which does not violate
diffeomorphism invariance \cite{Blen,OptrOrd,Bor}, but in the
planar case one may employ a simple point-splitting
regularization \cite{HSm}.  I arrange the three P's in the
$(\E{z}{Z})^{-1} $ term along the z
axis as follows.
\begin{equation}
     (\E{z}{Z}(z))^{-1} \epsilon _{AB} \A{A}{x}(z+ \epsilon)
\A{B}{y}(z-\epsilon) + (\epsilon \rta -\epsilon)
\label{eq:2.3c}
\end{equation}
As mentioned in the introduction, this term annihilates the
solutions $\psi $ constructed in the next section; consequently
the exact form of this
term matters little in the present paper, and a lengthy
discussion here about the details of regularization would be
pointless.  However, I do need to know enough about the action of
the $(\E{z}{Z}(z))^{-1}$ operator in this term to prove that this
operator is finite when acting on the solutions $\psi $.  As
shown in an
appendix, the \E{z}{Z} operator can be inverted, and proved to be
finite, provided the $S_Z $ SU(2) generator in the holonomy never
has a zero eigenvalue.  To see the reason for this restriction,
note that \EQ{2.2} plus \EQ{2.3a} imply
\begin{equation}
     \E{z}{Z}(z) \Theta (z_{i+1}, z,  z_i) M(z_{i+1}, z_i)
               = -i \hbar S_Z M \Theta (z_{i+1}, z, z_i).
\label{eq:2.3d}
\end{equation}
$\Theta $ is a double Heaviside or square wave operator
\begin{equation}
     \Theta (z_{i+1}, z, z_i) \equiv \theta (z_{i+1} - z)
                              \theta (z - z_i).
\label{eq:2.3e}
\end{equation}
If the $S_Z $ in equation~(\ref{eq:2.3d}) is zero, then \E{z}{Z}
has an identically zero eigenalue and cannot be inverted.
Therefore, I must assume $S_Z$ belongs to the (2j+1)x(2j+1)
dimensional representation, with j half-integer; or, if j is
integer, none of the $S_Z $ matrices in any holonomy takes on the
value zero.

     There is one other context in which the $(\E{z}{Z})^{-1}$
operator plays a role.  Suppose I wish to prove that the
constraint algebra closes, in the following restricted sense: if
I order the constraints on the right (in a typical constraint
commutator such as \EQ{1.5}) so that P's are to the right, Q's
are to the left in each constraint, and carry out the commutator,
then the constraints on the left in \EQ{1.5} are also ordered
correctly, with
P's to the right, Q's to the left, and operators such as $g^{ij}$
absent or to the left.  Although this is not full closure
(nothing is said about possible Schwinger terms), restricted
closure is a non-trivial result.  Teitelboim's rescaling allows
restricted closure.  Since the proof of this statement is
straightforward but lengthy, with the result presumably known
already to Teitelboim, I will not go through the entire proof,
but will mention only the step in the proof that is affected by
the presence of the $(\E{z}{Z})^{-1}$ operator.  In order to
commute this operator with \A{Z}{z} operators also present in the
constraints, I use
\begin{equation}
 [(\E{z}{Z} )^{-1},\A{Z}{z} ] = \hbar \delta (z-z')
                                   (\E{z}{Z})^{-2}.
\label{eq:2.3f}
\end{equation}
This is the naive commutator one obtains by sandwiching the
commutator of \EQ{1.2} between two factors of $(\E{z}{Z} )^{-1}$.

Although the double operator on the left in \EQ{2.3f} looks
dangerous, it is well-defined {\it on the solutions
studied in this paper} provided no $S_Z$ vanishes.  This is
enough to prove restricted closure valid when the constraints
act on the solutions of this paper.

     The fact that the constraints close, in the restricted
sense, with the operator ordering chosen, is an additional
confirmation that my earlier, tentative choice of operator
ordering is acceptable.  I mention that the constraint algebra
also closes for another popular choice of the P's and Q's: choose
the P's to be the connections, and order them to the right.

\section{Solutions away from Boundaries}

     This section proposes an ansatz for the solutions, then
verifies that the Hamiltonian $H_T $ annihilates these solutions
at all finite points of the z axis away from boundaries.
Strictly speaking it is not necessary to
prove $H_T\psi = H_T(\Np, N^z, N_G)\psi = 0$.  (In fact after the
surface terms are included,
$H_T\psi = $(ADM energy) $\psi$, which is non-zero.)  Rather, one
must prove $H_T(\delta
\Np, \delta N^z, \delta N_G)\psi = 0$, where the $\delta N$ are
arbitrary small {\it changes} in
the Lagrange multipliers N.  Small changes correspond to
infinitesimal changes in coordinates,
which should not make any physical difference.  In this section,
for simplicity of notation, I use
N rather than $\delta N$.  The distinction between N and $\delta
N$ becomes important at
boundaries, where a given N can approach a constant, forcing  the
corresponding $\delta N$
 to approach zero.

     I return to the Hamiltonian, equation~(\ref{eq:2.1a}) and
break it up into eigenstates of O(2) by writing out the
components of the Levi-Civita tensor,
\begin{equation}
     \epsilon _{-+} = -\epsilon _{+-} = i,
\label{eq:2.3b}
\end{equation}
while being careful to contract every + index with a - index, for
example $\epsilon _{MN} A_M B_N = \epsilon _{-+} A_{+} B_{-} +
\cdots $.  When this is done one finds that half of the terms in
$H$ contain at least one factor of $\A{-}{a} = \hbar \delta
/\delta \E{a}{+} $, hence half the terms would annihilate a
wave funtional containing only $\E{a}{-} $ fields.  I try
the ansatz
\begin{eqnarray}
     \psi (n;j) &=& \prod_{i=1}^n \int_{z_0}^{z_{n+1} } dz_i
          \theta (z_{i+1}-z_i) M(z_{n+1} ,z_n)\E{a_n}{-}(z_n)
          S_{+} \times  \nonumber \\
     & & \times M(z_n,z_{n-1})\cdots \E{a_1}{-}(z_1)
          S_{+}M(z_1,z_0 ).
\label{eq:2.4}
\end{eqnarray}
Since this ansatz contains no $\E{a}{+} $ fields, half the terms
in $H_T\psi $ vanish, including the term
involving $1/\E{z}{Z}$.  Put aside the $H_G$ term for the moment;
it will require special treatment.  The remaining terms all
contain ${\rm F}^{+}_{zb}$.

     Before evaluating the action of this operator on
$\psi $, note that each $\Etld (z_i)$ is integrated over $z_i$
(to preserve diffeomorphism invariance, as mentioned earlier),
and the limits of integration have been taken at two finite
but otherwise arbitrary points $z = z_0\ and\ z = z_{n+1}$,
rather
than $z = \pm \infty$.  As discussed in the introduction, nothing
is lost by considering an
arbitrary location for the surface.  In fact something is gained:
one no longer has to worry about
the convergence of the $dz_i$ integrals
in equation~(\ref{eq:2.4}).  Because of these limits of
integration, any theta function depending on $z_0\ or\ z_{n+1}$
in equation~(\ref{eq:2.4}) may be taken to be unity.

     If the matrices
$S_{+}$ in $\psi$ were the usual 2x2 Pauli lowering operators,
then $\psi $
would vanish identically (except for the n = 1 case).  I shall
take $S_{+}$ to be the $(2j+1)x(2j+1)$ representation, however,
so
that $\psi $ does not vanish identically, unless $n > 2j$.
Similarly, the $S_Z$ in the holonomy,
equation~(\ref{eq:2.2}), is $(2j+1)x(2j+1)$.

     The surviving terms in $(H_T-H_G)\psi $ are all proportional
to
${\rm F}^{+}_{zb}$:
 \begin{eqnarray*}
     (H_T-H_G)(z) \psi &=& -i(\Np + N ^z)\E{b}{-}
               {\rm F}^{+}_{zb}(z)\psi \\
          &\propto & \E{b}{-}[\partial _z(\delta \psi /\delta
\E{b}{-}) +i\A{Z}{z} \delta \psi /\delta \E{b}{-}] \\
          &=& \sum_{j=1}^n \int dz_n \cdots dz_j
     \theta (z_{j+1}-z_j) \theta (z_j-z_{j-1})\cdots dz_1
     M \cdots M\E{aj}{-}(z) \times \\   & &
\times[\partial_z + i\A{Z}{z} (z)]\delta (z_j - z)S_{-}M\cdots \\
          &=&\sum_{j=1}^n \cdots \delta (z_j - z) \E{aj}{-}(z)
               [\partial_{zj}+ i\A{Z}{z}] \times \\
     & &\times [\theta (z_{j+1}-z_j) \theta (z_j-z_{j-1}
               M(z_{j+1},z_j)S_{-}M(z_j,z_{j-1})] \cdots + IBP.
\end{eqnarray*}
IBP denotes  surface terms at $z_j = z_0\ and\ z_{n+1} $,
 which appear
when we turn $\partial_z \delta$ into $ -\partial_{zj} \delta$
and
integrate by parts.  All these surface terms are killed by the
theta functions, except one at  $z_{n} = z_{n+1} $ and one at
$z_1 = z_0 $.  I put the IBP term aside for the moment and
continue
working on the main term.  The $\partial_{zj} $ acting on the M's
produces a commutator $-i[S_z,S_{+}]\A{Z}{z} $ which just
cancels the $i\A{Z}{z} (z)$ term.  This leaves terms in which
$\partial _{zj} $ acts on the theta functions.  These terms are
\begin{eqnarray*}
     (H_T-H_G)(z)\psi & =& \sum_{j=1}^n \cdots
\E{aj+1}{-}(z_{j+1})
          M(z_{j+1},z_j)\delta (z_j - z) \E{aj}{-} (z) \times \\
     & \times& [-\delta (z_{j+1}-z_j) \theta (z_j-z_{j-1}) +
          \theta (z_{j+1}-z_j) \delta (z_j-z_{j-1})]\cdots + IBP,
\end{eqnarray*}
where delta functions with arguments depending on $z_{n+1}$ or
$z_0$
are understood to be zero, and theta functions with these
arguments are understood to be unity.   Then the (j+1)st term in
this
sum contains a $+\delta (z_{j+1}-z)\delta (z_{j+1} - z_j)$ term
which exactly cancels the $-\delta (z_j-z)\delta (z_{j+1} - z_j)$
coming from the jth term.  $(H_T-H_G)\psi $ then collapses to the
surface term IBP.
\begin{eqnarray}
(H_T-H_G)(z)\psi &=& i\hbar (\Np + N ^z)M(z_{n+1},z_{n+1})
          S_{+}\E{an}{-}(z_{n+1}) \delta (z - z_{n+1}) \times
                                             \nonumber \\
& & \times\int dz_{n-1}M(z_{n+1} ,z_{n-1})
          \cdots \int dz_1 \cdots M(z_1,z_0)- \nonumber \\
& &-i\hbar (\Np + N ^z)\int dz_n M(z_{n+1},z_n) \cdots \int dz_2
          M(z_2,z_0 ) \times  \nonumber \\
& & \times S_{+} \E{a1}{-}(z_0 ) M(z_0 ,z_0 )\delta (z - z_0 ).
\label{eq:2.5}
\end{eqnarray}
For general choice of the \Etld, $(H_T-H_G)(z)\psi $ is pure
surface term, non-zero only at the boundaries $z = z_0\ and\
z_{n+1}$.  Note the delta functions imply $(\Np + N ^z)$ should be
evaluated at the boundaries, and recall the remark at the beginning
of this section that the
Lagrange multipliers N  are really small changes $\delta N$.   In
the next section I shall require
both \Np\ and $N^z$ to approach constants at the boundaries;
therefore the corresponding $\delta
N$ vanish.  These surface terms will turn out to be harmless.

     The calculation of $H_G \psi $ is very similar to the
calculation just given. Cancellations occur at internal points,
but terms survive at the boundaries.  From
equations~(\ref{eq:2.2}) and (\ref{eq:2.3a}), the $\partial
_z\E{z}{Z} $ term in $H_G $ produces $S_Z $ terms:
\begin{equation}
    -i \partial _z\E{z}{Z} M(z_j,z_i) = \hbar [S_ZM \delta
(z-z_j) -MS_Z  \delta (z-z_i)].
\label{2.5a}
\end{equation}
As in the previous calculation, at points away from boundaries
the factors of $S_Z$ commute with $S_{+}$ and cancel the action
of
the other, $\epsilon_{IJ}\E{a}{I} \A{J}{a}$ term in $H_G$.
 At boundaries there is a surviving surface term:
\begin{equation}
     H_G(z) \psi = \hbar [N_G \delta (z-z_{n+1}) S_Z \psi -  N_G
\delta(z-z_0) \psi S_Z].
\label{eq:2.5b}
\end{equation}
This term would vanish if $N_G$ vanished at boundaries; but $N_G$
does not vanish (or go to a constant) in the planar symmetry
case, as the next section
will show.

     There is still the possibility that
the $H_G \psi $ term, \EQ{2.5b}, and the terms from $H_T - H_G$
will be canceled by the surface terms in the Hamiltonian, which
have the same support but have been omitted up to now.  When
surface terms $H_{st}$ are
included and the Hamiltonian density is integrated over
z, the constraint equation will have the following form.
\begin{eqnarray}
     0 &=& [\int_{z_0}^{z_{n+1}} dz H_T(z) + H_{st}]\psi
                                             \nonumber \\
       &=& \int dz [H_T(z) + \delta (z-z_0)\ and\ \delta
                         (z-z_{n+1})\ terms]\psi.
\label{eq:2.6}
\end{eqnarray}
The next section will
calculate $H_{st}$ , and section 5 will check for
cancellations.

     It is possible to construct an additional set of solutions
by replacing every $\E{a}{-} S_+$ in equation~(\ref{eq:2.4}) by
$\E{a}{+} S_-$.  I have not been able to mix both $\E{a}{-}$
and
$\E{a}{+}$ in the same wave functional, however.  The problem
comes from a
term proportional to $\Np [\E{b}{-} {\rm F}^{+}_{zb}-\E{b}{+}
{\rm
F}^{-}_{zb}]$ in $H_T$.  The minus sign between the two F's
destroys
the delicate cancellations previously obtained between
neighboring $z_j\ and\ z_{j+1}$ terms in
$\psi $. This lack of cancellation means that it is difficult to
implement procedure (4) in the
Introduction (procedure (4): search for closed flux solutions)
using a wave functional of the
present type (a string of $\E{a}{-}$ and $\E{a}{+}$ fields
separated by holonomies).
 Since a closed loop involves a trace, $\psi $ would
have to contain equal numbers of $\E{a}{-}$ and $\E{a}{+}$
fields (equal numbers of raising and lowering operators $S_+\
and\ S_-$).  As
suggested in the Introduction, implementing procedure (4)
presumably requires a radically new ansatz.

     Is it possible to insert an arbitrary function f($z_1\cdots
z_n$) into $\psi $?  In the linear case such a function (or its
Fourier transform) is present in the wave functional and
determines the spectral content of the wave packet.  Here,
however, one would obtain numerous $\partial_z f$ terms at the
integration by parts step, and $H_T \psi $ would no longer
vanish.  Note that not even $H_S \psi$ would vanish.  This is
signifigant because Husain and Smolin
\cite{HSm} propose j = $\frac{1}{2}$ f-dependent solutions  which
are
annihilated by $H_S$ (and $H_G$).
They then transform to loop space in order to satisfy the
remaining $H_z $ constraint.  This trick
does not work unless $H_S \psi$ vanishes.

\section{Surface Terms}

     The literature contains many discussions of
surface terms.   The
classical discussions based on metric dynamics \cite{deW,RTei},
have been updated recently to systems with the most
exotic boundary conditions \cite{HH}.  Within the framework of
connection dynamics
alone there are at least three recent discussions
\cite{Alect,Thiemann,BMP}.
Nevertheless, the planar case has enough twists and turns to make
another
discussion interesting, as well as necessary.

     The typical discussion of surface terms begins with two
assumptions.   The first assumption (algorithmic assumption)
justifies the need for the surface terms and
provides an algorithm for calculating these terms.  Since the
algorithm typically requires detailed knowledge of the behavior
of the basic fields on the boundary,
a second assumption (asymptotic assumption) is needed to supply
this behavior.

     I utilize an algorithmic assumption proposed by Regge and
Teitelboim \cite{RTei}.  When the Hamiltonian (more precisely,
the classical Lagrangian) is varied
to obtain the equations of
motion, $\delta $[$\partial_z$ (field)] variations occur which
must
be replaced by $\delta $(field)  variations.  The $\partial _z$
may
be removed by integration by parts, which in turn gives rise to a
surface term containing $\delta $(field).  The total variation
is required to
vanish (surface term as well as volume integral).  The vanishing
of the volume integral gives the classical equations of motion,
of course, but the
surface term does not vanish, in general, unless one adds a
compensating surface term to the original Hamiltonian.

     To see how the algorithm works in practice, let us apply
this method to the present Hamiltonian $H_T $,
equation~(\ref{eq:2.1a}).  Vary $H_T $ and look for $\delta$
[$\partial_z$ (field)] terms.  Every field strength F
conttibutes a term of the form  \Etld $\partial _z \delta A$.
Also, the Gauss constraint $H_G$ contributes a
$\partial _z \delta \E{Z}{z}$ term.  After integrating by parts
to remove $\partial _z $ from each $\delta $(field),
\bea
     \delta \int dz H_T &=& volume\ integral +
     [\Np \epsilon _{MN} \E{b}{M} \delta \A{N}{b} +
     iN^z \E{b}{M} \delta \A{M}{b} \nonumber \\
          & &-i\delta \E{Z}{z} \A{Z}{t}]\mid _{z_0 }^{z_{n+1}}.
\label{eq:3.1a}
\eea
I have used a result from the 3+1 decomposition, that the
Lagrange multiplier $N_G$ is also the four-dimensional
connection field \A{Z}{t} \cite{Alect,Rovlect}.
One can try to
cancel these surface terms by adding to the original
Hamiltonian $\int_{z_0}^{z_{n+1}} H_T$ a surface term $ H_{st}$
of the form
\begin{equation}
     \int dz H _T+ H_{st} = \int dz H_T - [\Np \epsilon _{MN}
          \E{b}{M} \A{N}{b} +
     iN^z \E{b}{M} \A{M}{b} ]\mid _{z_0 }^{z_{n+1}}.
\label{eq:3.1b}
\end{equation}

     At first glance this surface term does not seem to work.
Its variation does indeed give $\delta A$  terms
which exactly cancel the
corresponding $\delta A$ terms in equation~(\ref{eq:3.1a}); but
$ \delta H_{st}$ seems to be missing an \A{Z}{t} term needed to
cancel the corresponding \A{Z}{t} term in \EQ{3.1a};
and $ \delta H_{st}$ contains  $\delta \E{b}{M}\ ,\delta N^z$,
and $\delta \Np $ terms which are unwanted.   I could supply the
missing \A{Z}{t} term by adding to $H_{st}$ an additional term
$+i\E{z}{Z} \A{Z}{t} $ ; but again, this term generates an
unwanted $\delta \A{Z}{t}$ variation.  For now I will eschew
\A{Z}{t} terms.  (We shall see later that they are unnecessary;
\EQ{3.1b} is correct as it stands.)

     At this point I need the asymptotic
assumption to show that the "unwanted" $\delta $\E{b}{M}\
terms have exactly the right asymptotic behavior to cancel the
\A{Z}{t} term in equation~(\ref{eq:3.1a}), while the quantities
$\Np$ and $N^z$ approach
constant values which are not subject to variation at the
boundaries (hence $\delta N^z\ and\ \delta \Np $ vanish
there).  To formulate an asymptotic assumption, one falls
back on classical
experience and adopts for the quantum case the simplest
asymptotic
behavior that works in the classical case for the system of
interest.  In the case of a three-dimensional system with bounded
sources, for instance, the asymptotic assumption is flat space at
infinity.  In the one-dimensional radiative case, the analog of a
bounded source is a wave packet or packets, located inside
the region $z_0 < z < z_{n+1}$, and zero at the boundaries
($z_0,z_{n+1}$).   From Newtonian gravity it is too much to hope
that the
metric for this system is flat at the boundaries, but Szekeres
\cite{Sz} uses hypersurface orthogonality of the null tetrads
plus the field
equations to show that an assumption of {\it conformal}
flatness in the variables (z,t) is always possible .  To make
this idea precise, I introduce Szekeres' parameterization for
the plane gravitational wave metric.
\[
     ds^2 = e^A [dx^2 e^B  \cosh W + dy^2 e^{-B} \cosh W -
2dxdy \sinh W ]
\]
\be
     + e^{D-A/2} \{ [ -(\Np)^2  + (N ^z)^2]dt^2 +2 N^z dzdt
+ dz^2 \}.
\label{eq:3.2}
\ee
Conformal flatness at the boundaries means
\begin{equation}
     \Np \rta 1; N^z \rta 0,
\label{eq:3.3}
\end{equation}
so that the (z,t) sector of the metric assumes a $e^{D-A/2}[-
dt^2 + dz^2] =  -2e^{D-A/2}dudv$ form, flat except for a scale
factor.  Equation~(\ref{eq:3.3}) is already a considerable
simplification, since it means \Np\ does not have to be varied in
$H_{st}$, and all $N^z$ surface terms may be dropped.

     It should be possible to simplify the metric still further
at the boundary, since gravitational wave degrees of freedom
vanish there.  In the linearized limit, the fields B and W
defined in \EQ{3.3}
are amplitudes for the two polarizations of the gravitational
wave. This result suggests that in the exact theory B and W
should be taken as wave packets which vanish at the boundary.
\begin{equation}
     B \rta 0; W \rta 0.
\label{eq:3.4}
\end{equation}
Co nsistent with assumption~(\ref{eq:3.4}), the exact classical
equations of motion for B and W are hyperbolic.  (This result is
straightforward to prove but requires a digression into the
classical equations of motion and formulas from paper I, and I
relegate the proof to Appendix B.)

     Now only the fields exp(A) and D remain at the boundaries.
In the linearized limit, D = 0 and exp(A) = 1, while in quadratic
order exp(A) obeys a parabolic, rather than hyperbolic equation
with the linearized energy as source \cite{I}.
\[
   2(\exp A),_{zz} + [(B_z)^2 + (W_z)^2 + (\pi_B)^2 + (\pi_W)^2]
=
0.
\]
$\pi_X $ is the momentum conjugate to X.  Thus exp(A) is a good
candidate for long-range scalar potential,
but not for wave packet behavior.  D is gauge-sensitive and could
be anything, from the arguments given so far .

     However, D turns out to vanish at the boundaries.  The
argument was given in paper I, but is perhaps worth repeating
here, because it does not use the usual asymptotic assumption.
Paper I is based on geometrodynamics, so that there is no $H_G$
constraint, and the $H_S\ and\ H_z$ constraints are expressed in
terms of the fields A, B, D, W and their conjugate momenta.  To
determine D at the boundary, I
need only the part of $H_T$ which is independent of B, W,
and $N^z$, therefore does not vanish at the boundary:
\begin{eqnarray}
     \delta [\int dz H_T + H_{st}] &=& \delta \int \Np[2(\exp
A),_{zz} - (\exp A),_z D,_z + \cdots] + \delta H_{st} \nonumber
\\
     &=& volume\ terms + [2(\delta \exp A),_z - (\delta \exp A)
               D,_z \nonumber \\
     & & - (\exp A),_z \delta D + \delta H_{st}]
                              \mid_{z_0}^{z_{n+1}},
\label{eq:3.5}
\end{eqnarray}
where $\cdots$ denotes terms which vanish at the boundary or
contain no derivatives.  Obviously the +2($\delta \exp A),_z $
term can be canceled by inserting a $-2(\exp A),_z $ term into
$H_{st}$; but what term can I insert to cancel the $\delta \exp
A$
and $\delta D$ variations?  It turns out there is no such term,
and I must take
\begin{equation}
     D \rta 0
\label{eq:3.6}
\end{equation}
at boundaries, to eliminate variations that cannot be
canceled.

     To prove that there is no such term, one may try to
construct such a term until frustration sets in; or one may
construct a formal proof as follows.  Suppose such a term exists;
call it f.  The functional f = f[expA,D] occurs in $H_{st}$, and
its variation cancels the
D-dependent IBP terms coming from the volume term:
\[
     0 = - (\delta \exp A)D,_z - (\exp A),_z \delta D +
          \int dz' \{\delta e^A(z')[\delta f(z)/\delta e^A(z')]
\]
\[
               + \delta D(z')[\delta f(z)/\delta D(z')].
\]
This implies
\begin{eqnarray*}
     \delta f(z)/\delta e^A(z') &=& \delta (z-z') D_z; \\
     \delta f(z)/\delta D(z') &=& \delta (z-z') (e^A)_z.
\end{eqnarray*}
This is impossible because the two second functional derivatives
are not equal, QED:
\begin{eqnarray*}
     \delta^2 f(z)/\delta D(z'') \delta e^A(z') &=&
               \delta (z-z')\partial_z \delta(z-z'');\\
     \delta^2 f(z)/\delta e^A(z') \delta D(z'') &=&
               \delta (z-z'')\partial_z \delta(z-z').
\end{eqnarray*}

     The asymptotic behavior of all fields except D was
determined by the usual asymptotic assumption (which is: the
quantum fields have the same asymptotic behavior as the classical
fields).  D, on the other hand, was determined solely by the
algorithmic assumption, without invoking the asymptotic
assumption at all!  No cancelling $H_{st}$ term can be found;
therefore the Lagrangian formulation is not consistent unless D
is set equal to zero.

     So far I have used the algorithmic and asymptotic
assumptions to find the asymptotic behavior of the usual metric
fields (more precisely the fields B, W, $\cdots$ used by Szekeres
to parameterize the metric).  It is not a good idea to apply the
algorithmic and asymptotic assumptions directly to the Ashtekar
triad and connection fields, because their classical behavior is
poorly understood.  However, it is now straightforward to obtain
the asymptotic behavior of the tetrads, triads, and
connections, since these fields may be expressed in terms of the
metric fields.

     I start with the tetrads.  In the (x,y) sector the general
formulas for the tetrads as functions of B, W, $\cdots$ is quite
complicated, but the formulas are needed only at the boundaries,
where the metric is diagonal and the expressions simplify
considerably.
\begin{equation}
     e^{\pm}_a \rta \delta ^{\pm}_a \exp (A/2 \pm i \phi)
\label{eq:3.7}
\end{equation}
$\phi$, the angle of rotation around the Z axis, is the new field
which appears on switching variables from the 3-metric (four
independent metric fields) to the triads (five independent triad
fields).  The
classical behavior of this field is
\begin{equation}
     \phi(z,t) \rta constant,
\label{eq:3.8}
\end{equation}
so that $\delta \phi = 0$ at boundaries.  The expression for
$H_{st}$ is a scalar under internal Z rotations, hence cannot
depend on $\phi$.  Without loss of generality I may take $\phi
\rta 0$, therefore.

     The (z,t) sector of the
tetrad is simple enough that one can write down  expressions
true for all z, then take their limit as z \rta boundaries:
\begin{eqnarray}
     e^T_t &=& \Np \sqrt g_{zz} \rta \exp(-A/4);\label{eq:3.9} \\
     e^Z_z &=& \sqrt g_{zz} \rta \exp(-A/4);\label{eq:3.10}\\
     e^Z_t &=& N^z \sqrt g_{zz} \rta 0;\label{eq:3.11}\\
     e^T_z &=& 0.
\label{eq:3.12}
\end{eqnarray}
(These tetrads embody the standard choice of gauge for Lorentz
boosts along the Z axis; for the advantages of this choice,
equation~(\ref{eq:3.12}), see
Peldan \cite{Pel}.)

     Given these tetrads, one can now determine the \Etld\ fields
from the relation
\begin{equation}
     \E{i}{I} = e e^i_I .
\label{eq:3.13}
\end{equation}
The result is
\begin{eqnarray}
     \E{z}{Z} &=& \exp A; \label{eq:3.14} \\
     \E{a}{\pm} &\rta & \delta ^a_{\pm} \exp (A/4).
\label{eq:3.15}
\end{eqnarray}
Relation~(\ref{eq:3.14}) is correct for all z.

     Now one may compute the connections \A{I}{i} by computing
the four-dimensional Lorentz connection
$\omega^{IJ}_a$, then expressing the \A{I}{i} in terms of the
$\omega^{IJ}_a$.
\begin{equation}
     \omega ^{IJ}_a = [\Omega _{i[ja]} + \Omega _{j[ai]} -
               \Omega_{a[ij]}]e^{iI} e^{jJ},
\label{eq:3.16}
\end{equation}
where
\begin{equation}
     \Omega _{i[ja]} = e_{iM}[\partial _je^M_a -
                     \partial  _ae^M_j]/2.
\label{eq:3.17}
\end{equation}
(Since the fields $\omega\ and\ \Omega$ are four-dimensional, they
should carry a superscript (4) which I suppress for simplicity.)
Equation~(\ref{eq:3.16}) simplifies considerably at boundaries,
or wherever the tetrad matrix becomes diagonal.
\begin{equation}
     \omega ^{IJ}_a \rta [e_{aK}(\partial_j e^K_i -
               \partial_i e^K_j)]e^{iI} e^{jJ}.
\label{eq:3.18}
\end{equation}
Using these equations, one finds
\begin{eqnarray}
     \omega^{XY}_z &\rta & 0;\\
\label{eq:3.19}
     \omega^{TZ}_z &\rta & \exp (A/4)\partial_t \exp(-A/4);\\
\label{eq:3.20}
     \omega^{ZC}_a &\rta & -\delta^C_a \exp (-
          A/4)\partial_z (\exp A )/2;\\
\label{eq:3.21}
     \omega^{TB}_a &\rta & \delta^B_a \exp (-
     A/4)\partial_t (\exp A )/2;\\
\label{eq:3.22}
     \omega^{XY}_t &\rta & 0; \\
\label{eq:3.23a}
     \omega^{TZ}_t &\rta & \exp (A/4)\partial_z \exp(-A/4),
\label{eq:3.23b}
\end{eqnarray}
where $ B,C = \pm$ only.
These are the only Lorentz connection components needed to
calculate the \A{I}{i}.   The latter follow from the equations
given in Appendix A:
\begin{eqnarray}
     G\A{Z}{z} &=& - \omega^{XY}_z -i\omega^{TZ}_z \nonumber \\
               &\rta & -i\exp (A/4)\partial_t \exp(-A/4);
                                             \label{eq:3.24} \\
     G\A{B}{a} &=& -\epsilon_{BZC} \omega^{ZC}_a -i\omega^{TB}_a
                                             \nonumber \\
               &\rta & -\epsilon_{Ba} \exp (-A/4)
                         \partial_z (\exp A )/2  \nonumber \\
               & &  -i \delta^B_a \exp (- A/4) \partial_t
                         (\exp A )/2; \label{eq:3.25a} \\
     G\A{Z}{t} &=& -\omega^{XY}_t -i\omega^{TZ}_t \nonumber \\
                      &\rta & -i\exp (A/4)\partial_z \exp(-A/4).
\label{eq:3.25b}
\end{eqnarray}
Note the last equation: in contrast to the situation in three
dimensions, the Lagrange multiplier for the Gauss constraint does
not go to zero at the boundaries.

     Since the asymptotic behavior of the triads and connections
depends on the asymptotic behavior of the $\exp A $ field, I take
a moment to investigate the asymptotic behavior of the latter.
It is relatively easy to extract this behavior from the
geometrodynamical equations derived in paper I.
After a canonical transformation
to the A,D,B,W parameterization, the Lagrangian assumes the form
\begin{eqnarray}
     L &=& \pi_A A_t + \pi_D D_t + \cdots - H_T \nonumber \\
       &=& \pi_A A,_t + \pi_D D,_t + \cdots \nonumber \\
       & & - \Np[2(e^A),_{zz} -
               (e^A),_z D,_z -e^{-A}\pi_A \pi_D +\cdots ],
\label{eq:3.26}
\end{eqnarray}
where $\pi_X $ is the momentum conjugate to X, $\cdots $ denotes
terms depending on B, W, and $N^z$ (which do not contribute near
boundaries) as well as the $H_{st}$ term (which may be ignored
since its
variation does not give the classical equations of motion).
Setting the scalar constraint equal to zero gives a parabolic
equation for $e^A$
which can be solved provided the $e^{-A}\pi_A \pi_D $ term is
known.  From the variation of the Lagrangian with respect to
$\pi_D $,
\[
     D,_t +\Np e^{-A}\pi_A = 0.
\]
Hence from equation~(\ref{eq:3.6}) for D, $\pi_A $ vanishes at
the boundary.  When this result
is inserted into the scalar constraint, $(\exp A),_{zz}$
is found to vanish; consequently $ \exp A $ at boundaries behaves
like the Newtonian
scalar potential in one dimension.
\begin{equation}
     \exp A \rta c_1z + c_2,
\label{eq:3.27}
 \end{equation}
where $c_1\ and\ c_2 $ are time-dependent functions. Moreover,
2$c_1$ is the ADM energy \cite{I}, therefore does not vanish, in
general.

     I now have the asymptotic behavior of all fields which occur
in the surface terms, and can return to the question raised at
equation~(\ref{eq:2.6}): do the surface terms from $H_T $,
equation~{\ref{eq:3.1a}), cancel the surface terms from the
variation of $H_{st}$, equation~(\ref{eq:3.1b}) ?  The total
surface term  obtained on variation is
\begin{eqnarray}
     \delta \int dz H_T + \delta H_{st} &=& volume\ term +
          [-i\delta \E{Z}{z} \A{Z}{t}]\mid _{z_0 }^{z_{n+1}}
          - [\Np \epsilon _{MN} \delta \E{b}{M} \A{N}{b} ]
                         \mid_{z_0 }^{z_{n+1}} \nonumber \\
          &\rta & volume\ term + [-i\delta (e^A)(i)(A/4),_z]
                                             \nonumber \\
          & & -[\epsilon _{MN} \delta (\delta ^b_{M} \exp (A/4))
             (-1)\epsilon_{Nb} \exp (-A/4)\partial_z (\exp A)/2]
                              \nonumber \\
          &=& volume\ term + [e^A \delta A A,_z/4] -
               [e^A\delta A A,_z/4 ]
                               \nonumber \\
          &=& volume\ term.
\label{eq:3.28}
\end{eqnarray}
On the first line the first square bracket comes from $H_T$ and
the second from $H_{st}$.   $N^z \rta 0$ terms have been
dropped; and the $\delta $\A{N}{b} terms from the two brackets
have canceled out.  On the second line, the Kronecker
$\delta^N_b$ part of
\A{N}{b}, \EQ{3.25a}, has been dropped; it cancels in a later
sum over indices.  Note in the first bracket
the \A{Z}{t} Gauss term not
only survives to infinity; it is actually needed at infinity, to
cancel the $\delta \E{b}{M}$ term from the second bracket.  From
the last line of \EQ{3.28} all surface terms have canceled from
the Euler-Lagrange variation, QED.

     I have also experimented with a more complex $H_{st}$ which
includes an \A{Z}{t} term:
\[
     H_{st}' = a_1 H_{st} + a_2 i\E{z}{Z} \A{Z}{t},
\]
where $a_1\  and\ a_2$ are adjustable constants.  When varied,
the new surface term generates not only $e^A \delta A A,_z$
terms, of the same form as \EQ{3.28}, but also new terms of the
form $\delta (e^A),_z$.  Requiring both types of variation to
vanish independently places two constraints on the two constants
$a_i$, and the only solution is $a_1 = 1, a_2 = 0$.  There are no
\A{Z}{t} terms in $H_{st}$

\section{Solutions at Boundaries; Classical Matter}

     The surface terms $H_{st}$ derived in section 4,
equation~(\ref{eq:3.1b}), can now be used to answer the question
raised in section 3: does $H_{st} $ cancel the integration by
parts boundary terms obtained when $H_T $ is applied to $\psi $?
The expression which must vanish is
\begin{equation}
     (\int dz H_T + H_{st})\psi.
\label{eq:4.1}
\end{equation}
More precisely, $\psi $ must be invariant under infinitesimal
transformations generated by the Hamiltonian.  If $H_T = H_T[\Np,
N^z, N_G]$, and similarly for $H_{st}$, then the expression which
must vanish is
\begin{equation}
     (\int dz H_T[\delta \Np, \delta N^z, \delta N_G] +
          H_{st}[0, 0,\delta N_G ]\psi,
\label{eq:4.2}
\end{equation}
and it must vanish for arbitrary choice of the small quantities
$\delta N $.  Note the zero arguments in equation~(\ref{eq:4.2}):
two of the N's are fixed at the boundary by
equations~(\ref{eq:3.3}), and only $N_G \equiv \A{Z}{t} $ can be
varied at the boundary.  However, from equation~(\ref{eq:3.1b})
there are no $N_G$ or \A{Z}{t}  terms in $H_{st} $.  The surface
term in  expression~(\ref{eq:4.2}) is identically zero.  The
integration by parts surface terms from $H_T $ remain uncanceled.

     The introductory section suggested several ways to modify
the Hamiltonian or wave functional so as to eliminate the surface
terms and obtain a solution to
the constraints.  The remainder of this section will
implement procedure (2) suggested in the Introduction: add
approximate Fermionic matter terms to the Hamiltonian and include
the Fermion fields in the wavefunctional, so that the flux lines
terminate on Fermion fields and the modified wave functional is
annihilated by the Gauss constraint.  In order to describe
quantitatively how much flux exits at each boundary, I introduce
some notation.  Let $m_i$ be the $S_Z$ eigenvalue of the homotopy
matrix to the left of $\E{b}{-}(z_i) S_+$ in $\psi$.
When previously suppressed matrix indices are exhibited,
\bea
     (\psi)^{m_n}_{\ \ m_0} &=& [M(z_{n+1},z_n)\E{an}{-}(z_n) S_+
\cdots \E{a1}{-}(z_1) S_+ M(z_1,z_0)]^{m_n}_{\ \ m_0} \nonumber
\\
    &=& M^{m_n}_{\ \ m_n} \E{an}{-} (S_+)^{m_n}_{\ \ m_{n-1}}
       \cdots \E{a1}{-} (S_+)^{m_1}_{\ \ m_0} M^{m_0}_{\ \ m_0}.
\label{eq:4.3}
\eea
Evidently the matter at $z_{n+1}$ must have $S_Z = m_n$, while
the matter at $z_0$ must have $S_Z = m_0$.

     The added Fermionic matter terms may be either quantized
(operator fields) or classical (c-number functions).  Suppose
first the matter terms are quantized; take them to be (say) the
Hamiltonian for a free spin 1/2 Weyl Fermion.  "Free" means that
covariant derivatives are replaced by ordinary derivatives and
spatial tetrads are replaced by unity, but Lagrange
multipliers are left unchanged, so that the matter terms appear
as free-field additions to the usual scalar, vector, and Gauss
constraints.  Then the algebra of constraints will continue to
hold.  (Note that the structure constants of this algebra are
independent of the gravity fields, because of the rescaling
suggested by Teitelboim; hence the interacting Fermion and free
Fermion terms obey the same algebra.)  One can make $\psi$ into a
Gauss scalar by sandwiching $\psi$ between initial and final
Fermion wavefunctions.  These are constructed by multiplying
together
enough spin 1/2 Fermion fields to get the required total $S_Z$ of
$m_0 \ or\ m_{n+1}$.  (At this point each Fermion in the product
should be given a distinguishing "color" index, to avoid
difficulties with Fermi statistics.)  If the spin 1/2 Fermi
coordinates are used to construct these wavefunctions, rather
than the
canonical Fermi momenta, then these initial and final
wavefunctions are not densities and do not have to be integrated
over $z_0\ and\ z_{n+1}$.  This construction is
straightforward enough, but for consistency I would have to
multiply the gravitational wave functional by a wave functional
for the Fermi field (or rather by a product of such wave
functionals, one for each color).  This seems unnecessarily
elaborate.

     I therefore turn to classical matter.  Here there is no hope
of getting a closed algebra of constraints, but closure of the
matter terms is not required for consistent quantization anyway.
Consider for example the scalar-scalar commutator, which produces
a linear combination of the vector and Gauss constraints.
Schematically,
\[
     [H_S.H_S] = H_z + H_G.
\]
Even if the scalar constraints on the left have c-number matter
terms, they certainly will not survive commutation and appear on
the right.  However, recall that these commutators are smeared
with small changes $\delta \Np_1\  and\ \delta \Np_2$.  These
changes vanish at the boundaries z = $z_b$, hence  if I take the
c-number corrections to be proportional to $\delta(z-z_b)$, they
will disappear from the
commutator algebra.  (The $[H_G,H_G]$ commutator does
not give trouble, even though the smearing function $\delta
\A{Z}{t}$ is finite at boundaries, since this commutator vanishes
identically.)  I therefore propose the following c-number
modification of the Gauss constraint.
\be
     H_G \rta H_G + \hbar m_0 \delta(z-z_0)
                         -\hbar m_n \delta(z-z_{n+1}).
\label{eq:4.4}
\ee
There are analogous, delta-function modifications of the $H_S\
and\ H_z$ conmstraints, but their explicit form will not be
needed.

     I must now return to sections 2 through 4 of this
paper, and check that all boundary conditions and gauge choices
adopted in those sections continue to hold in the presence of the
matter.  In section 2 ("Quantization") I eliminated four
constraints by adopting the gauge due to Husain and Smolin
\cite{HSm}.  I must check that this gauge choice still eliminates
these
constraints (the
$H_x,\ H_y $ diffeomorphism constraints and the $H_X,\ H_Y$ Gauss
constraints) when Fermionic matter is present.  Although I do not
quantize the Fermion matter, for orientation it will be helpful
to have at hand the Hamiltonian for the free spin 1/2 Weyl
Fermion.
\be
     H_F = - \Np \omega \partial_a \sigma ^I \eta
                    -N^a \omega \partial_a \eta
                    +i \A{I}{t}\omega \sigma ^I \eta /2 .
\label{eq:4.5}
\ee
$\eta^A$ is a spin 1/2 Weyl Fermion and $\omega_B$ is its
conjugate momentum.  In this section I reserve Roman
letters A,B, $\cdots$ from the beginning of the alphabet for spin
indices (A,B,$\cdots = \pm$ 1/2), and continue to use Roman
capitals I,J,$\cdots$ from the middle of the alphabet for the
Lorentz indices (I,J,$\cdots$ = X,Y, or Z).  Thus $\sigma^I$ is
shorthand for the Pauli matrix $(\sigma_I)^B_{\ A}$.  I note that
the diffeomorphism constraints $H_a$ with a = (x,y) in \EQ{4.5}
contain $\partial_a$.  Since this vanishes because of the
symmetry
generated by the spacelike Killing vectors, there are no matter
corrections to $H_x$ or $H_y$ in \EQ{4.5}, and I will assume that
there are likewise no c-number matter corrections to $H_x$ or
$H_y$ in my model.  In \EQ{4.5} the $H_X$ and $H_Y$ constraints
contain $\sigma ^I$, I = (X,Y).  If $m_0$ is positive (say), I
can build the initial Fermion wavefunction  entirely out of spin
1/2 Fermions having spin +1/2, while imposing the constraint that
the spin -1/2 $\eta$ and $\omega$ fields vanish.  Then the
$\sigma ^I$ terms for I = (X,Y) are constrained to vanish and
there are no matter corrections to $H_X$ or $H_Y$.  Taking my cue
from this example, I assume that the c-number Fermions used to
construct the initial wavefunction in my model are entirely
polarized in the $m_0$ direction, and similarly for the Fermions
in the final wavefunction, so that there are no c-number matter
corrections to $H_X$ or $H_Y$.

     Husain and Smolin fix the x and y diffeomorphism
gauges and the X and Y Gauss constraints by choosing
\begin{equation}
     \E{a}{Z} \approx \E{z}{I} \approx 0,
\label{eq:4.6}
\end{equation}
where a = x,y, I = X,Y, and $\approx$ as usual denotes a gauge
condition which is to be imposed only after carrying out all
commutators.  Now if one commutes the Hamiltonian (including
matter terms) with the constraints of \EQ{4.6}, one finds that
these constraints are conserved in time only if further
constraints hold:
\bea
     \A{Z}{a} &\approx& \A{I}{z} \approx 0; \label{eq:4.7} \\
     \A{I}{t} &\approx& N^a \approx 0. \label{eq:4.8}
\eea
These are exactly the additional constraints obtained by Husain
and Smolin, as one would expect, since the added matter terms are
c-number and do not change the equations of motion for the tetrad
and connection fields.  Further commutation
of the constraints~(\ref{eq:4.7}) with the Hamiltonian yields no
new constraint.

     At this point I can treat the constraints in \EQ{4.6} and
\EQ{4.7} as four pairs of second-class constraints, and eliminate
them using the Dirac bracket procedure.  Since the four pairs are
just canonical coordinate-momentum pairs, the "Dirac brackets"
for the remaining \Etld\ and A fields are just the usual Poisson
brackets for these fields.  After setting the constraints in
\EQ{4.6} and \EQ{4.7} strongly equal to zero, I reproduce the
Smolin-Husain result that the
gravitational part of the $H_a$ and $H^I$ constraints are
strongly equal to zero, for a = x,y and I = X,Y.  Since the
Fermionic part of these constraints also vanishes, all four of
these constraints vanish strongly, and I may set $N^a = \A{I}{t}
= 0$ for a = (x,y) and I = (X,Y): I no longer need the
equations obtained by varying $N^a$ and \A{I}{t} in the
Lagrangian. At this point I have eliminated four constraints and
four coordinate-momentum pairs, the same number as Husain and
Smolin. Their gauge choice continues to work in the presence of
c-number matter terms.

     In section 3 ("Solutions") it is easily verified that the
modified $H_G$ of \EQ{4.4} now annihilates the solution $\psi$ of
section 3: compare \EQ{4.4} to \EQ{2.5b}.  In section 4 ("Surface
Terms") I must review
Szekeres' argument \cite{Sz} that the metric may be brought to
conformally flat form
at infinity.  Szekeres uses the
Newman-Penrose spin coefficient formalism and shows that a
certain spin coefficient $\alpha$ must vanish.  From this it
follows that the metric decomposes into two disconnected 2x2
subblocks, a (z,t) subblock and an (x,y) subblock.  Then he
uses the existence of two hypersurface-orthogonal null vectors to
bring the (z,t) subblock (globally) to a conformally flat form.
The Newman-Penrose $\alpha$ happens to be a combination of the
connections which appear in \EQ{4.7}.  Therefore the Szekeres
argument remains valid, because the Husain-Smolin gauge remains
valid.

     Since the added matter terms are being treated as c-number
"background" matter, they are not to be varied when obtaining the
equations of motion; consequently, they contribute no integration
by parts surface terms; no matter terms need be added to
$H_{st}$.  The argument that $D\rta 0$ at  the boundary therefore
continues to be valid.  Only the argument that exp(A) is linear
at the boundaries, \EQ{3.27}, must be modified, since the scalar
constraint now contains matter terms.  All other results in
section 4 continue to hold.  This completes the check  that the
boundary conditions and gauge choices of sections 2-4 continue to
hold in the presence of c-number matter.

     I comment briefly on the effect of including interactions in
the matter terms.  For example, adopt the quantum operator
Hamiltonian of \EQ{4.5} with ordinary derivatives replaced by
covariant derivatives and with tetrad fields restored.  Because
of
the covariant derivatives, the Fermion Hamiltonian now contains
\A{J}{a} connection
terms.  If one tries to impose the gauge choice of Husain and
Smolin, \EQ{4.6}, the secondary
constraints of \EQ{4.7} are replaced by constraints of the form
\A{J}{a} $\approx$ (bilinear combination of Fermion fields).
This is a manifestation of the fact that in a theory with
minimally coupled Fermions and no "curvature squared" terms, the
gravitational connection acquires a torsion, with the torsion
non-dynamical and bilinear in the Fermion fields.  Because of the
torsion, the Newman-Penrose $\alpha$ no longer vanishes.
Evidently in a
theory with interacting Fermions, even the kinematics (gauge
choices) would have to be rethought from the beginning.

\section{Directions for Future Research}

     The Rovelli-Smolin symmetric states \cite{symmsta}, or
spin-network states
$\Gamma[\gamma]$ are functionals of loops.  The loops $\gamma$
are flux lines
of spin j which join and separate at trivalent vertices; the
vertices are Clebsch-Gordan coefficients.  Symmetric states and
the states $\psi[\Etld,A]$
introduced in the present paper obviously resemble each other,
but the exact relation between the two is not straightforward to
establish.  In three dimensions a functional Fourier transform
connects $\Gamma[\gamma]$
to $\psi[\Etld,A]$.  This transform was established by demanding
that certain $T^n$
operators defined in $(A,\Etld)$ space have simple analogs in
loop space \cite{RovSmo}.  When three dimensions are replaced by
one, the $T^n$ algebra changes dramatically, since most of the
$\A{A}{a} $ are no longer
connections, and the Fourier transform must be rethought
from the beginning.  This is beyond the scope of the present
paper.

     However, the Rovelli-Smolin construction does suggest  the
following generalization of the solutions of section 3.  In
$\psi$ the (2j+1)x(2j+1) $S_+$  raising operator multiplying each
$\E{a}{-}$ is proportional to a Clabsch-Gordan coefficient
\begin{equation}
     (S_K)_{m'm} = -\sqrt{j(j+1)} <jm'\mid 1Kjm>.
\label{eq:5.1}
\end{equation}
(Since the coefficient on the right obeys Condon-Shortley
conventions \cite{Ed},
the $S_K$ on the left must be normalized like a Condon-Shortley
vector operator, which means $S_{\pm 1} = \mp(S_X \pm
iS_Y)/\sqrt{2}$.)
One can generalize equation~(\ref{eq:5.1}) to
\begin{equation}
     (S_K)_{m'm}\rightarrow  (const.)<j'm'\mid 1Kjm>,
\label{eq:5.2}
\end{equation}
where $j'$ may be $j \pm 1$ as well as j.   Simultaneously,
the (2j+1) dimensional holonomy M to the left of $S_+\E{a}{-}$ must
be replaced by a ($2j'+1$)
dimensional M.  Note the only property of the $S_K$ matrix used
in deriving the solutions of section 3 is
\begin{equation}
     [S_Z,S_K] = K S_K.
\label{eq:5.3}
\end{equation}
Written in terms of Clebsch-Gordan coefficients, this is an
obvious identity which generalizes immediately from
equation~(\ref{eq:5.1}) to equation~(\ref{eq:5.2}), so that the
matrices of equation~(\ref{eq:5.2}) also generate solutions.

     Rovelli and Smolin have constructed a gauge-invariant,
diffeomorphism-invariant, and regulated area operator.  I have
applied this operator to the solutions of the present paper, with
puzzling results.  This work is described in Appendix D.

     Husain and Smolin have constructed a large number of
solutions which are more general than the solutions presented
here, in the sense that the wave functional involves both $S_+$
and $S_- $ \cite{HSm}.  These solutions use only j = 1/2  and
satisfy the constraints because of
the special properties of the Pauli matrices.  Some (very)
preliminary work indicates that these solutions cannot be
generalized to the case $j > 1/2$.  The j = 1/2 sector may be the
best place to look for a lowest energy state, however, and the
Husain-Smolin
solutions deserve further study.

\appendix
\section{3+1 Phases}

      When the four-dimensional covariant
formalism (with real Lorentz connection $^{(4)}\omega^{IJ}_a)$ is
rewritten as a 3+1
dimensional canonical formalism (with complex connection
\A{I}{i}), one must make three choices of phase.  Two of these
choices are straightforward and require little discussion: choose
the phase of the Levi-Civita tensor by choosing the sign of
$\epsilon_{TXYZ} $; and
choose between self-dual or anti-self dual complex
connection by choosing a phase $\delta = \pm 1 $.
\begin{equation}
     2G ^{(4)}A^{IJ}_a = ^{(4)}\omega^{IJ}_a
     +i\delta (\epsilon^{IJ}_{\ \ KL}/2\epsilon_{TXYZ} )
^{(4)}\omega
                                   ^{KL}_a.
\label{eq:A1}
\end{equation}
The duality relation is
\begin{equation}
     i\delta (\epsilon^{IJ}_{\ \ KL}/2\epsilon_{TXYZ} )
^{(4)}A^{KL}_a
               = ^{(4)}A^{IJ}_a
\label{eq:A2}
\end{equation}
The explicit factor of $\epsilon_{TXYZ}$ simplifies later formulas
and guarantees that $\delta$ is
independent of choice of phase for the Levi-Civita tensor.

     The third choice of phase occurs when the connection
$^{(4)}A^{IJ}_a$ with two space-time Lorentz indices is replaced
by the connection  \A{I}{i} with a single spatial Lorentz index.
For
Lorentz indices M,N = space-space, the equation connecting the
one- and two-index connections is
\begin{equation}
     ^{(4)}A^{MN}_a = \sigma \epsilon_{MNS} \A{S}{a} /2,
\label{eq:A3}
\end{equation}
where $\sigma = \pm 1$ is the arbitrary choice of phase.  For
Lorentz indices T,M = time-space, the equation connecting one-and
two-index connections
follows from
equation~(\ref{eq:A3}) plus duality, equation~(\ref{eq:A2}).
\begin{equation}
     ^{(4)}A^{TM}_a = -i\sigma \delta \A{M}{a} /2 .
\label{eq:A4}
\end{equation}
The standard proof that the Ashtekar formulation of gravity is
equivalent to the usual formulation \cite{Lagr,Rovlect}
starts from a gravitational lagrangian L expressed in terms of
the real connection $^{(4)}\omega$; then a pure imaginary
topological term is introduced, so that L can be rewritten as a
function of the complex connection $^{(4)}A$; then
equations~(\ref{eq:A3})-(\ref{eq:A4}) are used to eliminate
$^{(4)}A$ and bring the Lagrangian to $p\dot{q} - H_T $ canonical
form.  The dependence of the final result on the above phase
conventions $\epsilon_{TXYZ}, \delta, and\ \sigma $ is as follows:
\begin{eqnarray}
     8\pi L = &-& i\sigma \delta \E{b}{B} \A{B}{b},_t
        + \ut{N} \sigma \epsilon_{MNS} \E{m}{M} \E{n}{N} F^S_{mn}/2
               \nonumber \\
       &+& i\sigma \delta N^m \E{n}{N} F^N_{mn}
        + i\sigma \delta \E{n}{N} D_n \A{N}{t}.
\label{eq:A5}
\end{eqnarray}
The phase $\epsilon_{TXYZ} $ has completely dropped out of the
final result.  One can read off the canonical momentum from the
coefficient of $\A{B}{b},_t $.  Since the momentum depends on
$\sigma \delta $, so does the equal-time commutator after
quantization:
\begin{equation}
     [\A{M}{m} , \E{N}{n}] = -\sigma \delta \hbar \delta^M_N
                    \delta^n_m \delta (z-z').
\label{eq:A6}
\end{equation}
F and $D_n$ in equation~(\ref{eq:A5}) are the three-dimensional
field strength and covariant derivative respectively.
\begin{eqnarray}
     F^S_{mn} &=& \partial_m \A{S}{n} - \partial_n \A{S}{m}
                    -\sigma \epsilon_{SMN} \A{M}{m} \A{N}{n} G;
                    \nonumber \\
     D_n \A{N}{t} &=& \partial_n \A{N}{t} - \sigma \epsilon_{NMR}
                    \A{M}{n} \A{R}{t} G.
\label{eq:A7}
\end{eqnarray}

     So far there is zero reason to prefer any particular phase
for $\epsilon_{TXYZ}$; and little or no reason to prefer any
particular phase for $\delta $.  (I have chosen +1 for both these
phases.)  What about $\sigma $?  The usual choice in the
literature appears to be +1 \cite{Rovlect}.  However, note that
traditional conventions for the matrix elements of the 3x3 J = 1
SU(2) generators $S_S $ imply
\[
     \A{S}{a} [S_S]_{MN} = +\epsilon_{MSN} \A{S}{a} ,
\label{eq:A8}
\]
with S in the middle rather than at the end, which suggests
$\sigma = -1 $ in equation~(\ref{eq:A3}).  If $\sigma = +1$ is
actually the unconventional choice for this phase, one should get
unconventional results in equations which are non-linear in the
fields \A{S}{a}.  Indeed, consider equations~(\ref{eq:A7}).  With
$\sigma = +1$, these are {\it not} the traditional
definitions of field strength and covariant derivative in SU(2)
gauge theories.  In this appendix I have left $\sigma\ and\ \delta
$ as arbitrary phases in the final formulas, so that authors may
make their own choice, but in the body of
the paper I have chosen  $\epsilon_{TXYZ}= \delta = +1,\ \sigma =
-1$ .

\section{Equations of Motion for the B and W Fields}

     This appendix verifies the assertion made at
equation~(\ref{eq:3.4}) that the classical equations of motion for
the B and W fields are hyperbolic.   From paper I, the Lagrangian
written in terms of Szekeres fields B, W, D, A is
\bea
     L& =& \pi _B \cosh W B,_t + \pi _W W,_t
          -\Np e^A [(B,_z \cosh W)^2 + (W,_z)^2]/2 \nonumber \\
        & &  - \Np e^{-A} [(\pi _B)^2 + (\pi _W)^2]/2
          -N^z [\pi _B \cosh W B,_z + \pi _W W,_z] + \cdots,
\label{eq:B1}
\eea
where $\cdots $ indicates irrelevant terms depending on A and D
and their conjugate momenta only.  I will prove the assertion for
the B field.  The proof for the W field proceeds similarly and is
slightly easier, because the equations contain no $\cosh W$
factors.

     The variation of L with respect to $\pi _B$ and B gives
\begin{eqnarray}
     0 &=& \cosh W B,_t - \Np e^{-A} \pi _B -N^z \cosh W B,_z;
                         \nonumber \\
     0 &=& -(\pi _B \cosh W),_t + [\Np e^A B,_z (\cosh W)^2],_z
\nonumber \\
         & &                + N^z [\pi _B \cosh W],_z.
\label{eq:B2}
\end{eqnarray}
One can obtain a wave equation for B in the usual way, by solving
the
first of these equations for $\pi_B $ and inserting the solution
into the second equation.
\[
     0 = -[e^A (1/\Np)( B,_t -N^z B,_z)(\cosh W)^2],_t
               + \Np e^A B,_{zz} (\cosh W)^2
\]
\be
     + N^z [e^A (1/\Np)  (B,_t -N^z B,_z)(\cosh W)^2],_z + \cdots,
\label{eq:B3}
\ee
where $\cdots $ are terms containing no second derivatives of B.
Note that to establish the hyperbolic character of wave \EQ{B3},
one needs only the second derivative terms \cite{CourHil}.
Carring out the differentiations in \EQ{B3}, keeping only the
second derivatives, one gets
\begin{eqnarray}
     0 &=& -e^A (1/\Np)(B,_{tt} -N^z B,_{zt})(\cosh W)^2
               + \Np e^A B,_{zz} (\cosh W)^2 \nonumber \\
       & &+ N^z e^A (1/\Np)  (B,_{tz} -N^z B,_{zz})(\cosh W)^2 +
\cdots
                         \nonumber \\
       &=& e^A (\cosh W)^2 (1/\Np)\{-B,_{tt} + 2 N^z B,_{zt} +
          [(\Np)^2 - (N^z)^2]B,_{zz}\} + \cdots \nonumber \\
      &\equiv & aB,_{tt} + bB,_{zt} + cB,_{zz} + \cdots.
\label{eq:B4}
\end{eqnarray}
The condition for a hyperbolic equation is $b^2 - 4ac > 0 $
\cite{CourHil}, which translates to $[e^A (\cosh W)^2]^2 >
0$, QED.  Note that it is necessary to divide through by \Np\ and
$e^A $ at several points in the above proof.  If these quantities
vanish, the metric of  \EQ{3.2} becomes degenerate, since the
determinants of the (z,t) or (x,y) subblocks vanish.

\section{Inversion of \E{z}{Z} }

     This appendix computes the action of
1/\E{z}{Z} on enough functionals to show that
1/\E{z}{Z} is bounded when acting on the solutions $\psi $ of
section 3.

     To invert an operator, such as the momentum operator
$\hat{p} $ in quantum mechanics, one might try the following
(naive) procedure.  Expand an arbitrary state in a complete set
of momentum eigenstates.  On each eigenstate the action of
$\hat{p} $ gives a constant,
\begin{equation}
     \hat{p} \exp (ikx) = \hbar k \exp (ikx).
\label{eq:C1}
\end{equation}
Then define the inverse as one over the constant:
\begin{equation}
     [1/\hat{p}] \exp (ikx) = [1/\hbar k] \exp (ikx).
\label{eq:C2}
\end{equation}
This definition has the advantage that $\hat{p} [1/\hat{p}]$
gives unity, as it should.  The problem is that k can vanish.

     When the spectrum of the operator is continuous (as is the
case for $\hat{p} $) one can try defining 1/k at k = 0, say by
replacing $k \rta k \pm i \epsilon $.  There is no point in
discussing this option here, since the spectrum of \E{z}{Z}
appears to be discrete: from \EQ{2.3b} its action upon a
holonomy gives a factor of $S_Z $.  Therefore vanishing
eigenvalues are a problem, and the conclusion given in section 2
appears unavoidable: holonomies with $S_Z = 0$ must be excluded.
Once the kernel of
\E{z}{Z} is excluded,  \E{z}{Z}
can be inverted using the method sketched above for $\hat{p} $:
find the eigenfunctions, and define the inverse as one over the
eigenvalue.

     Strictly speaking there is not one operator \E{z}{Z} (z) to
invert, but an infinite number, one for each value of z.  The
eigenvectors also depend on z, if the operators depend on z:
$\hat{O}(z) \phi (z) = \lambda (z) \phi (z) $.  The holonomy
times a $\Theta$ function is a z-dependent eigenvector:
\begin{equation}
     \E{z}{Z}(z) [\Theta (z_{i+1}, z, z_i) M(z_{i+1}, z_i)]
               = -i \hbar S_Z  [\Theta (z_{i+1}, z, z_i) M]
\label{eq:C3}
\end{equation}
$\Theta $ is the square wave function defined at \EQ{2.3e}.  The
holonomy M is defined at \EQ{2.2}, and I have used \EQ{2.3a}.
The inverse of \EQ{C3} is
\begin{equation}
     [1/\E{z}{Z}(z)] [\Theta (z_{i+1}, z, z_i) M(z_{i+1}, z_i)]
               = i(\hbar S_Z)^{-1} [\Theta (z_{i+1}, z, z_i) M]
\label{eq:C4}
\end{equation}

     Next consider the action of 1/\E{z}{Z} on the solutions
$\psi $ of section 3. These solutions have the form

\begin{equation}
     \psi = M(z_{n+1}, z_n) X(z_n) \cdots M(z_2, z_1) X(z_1)
                    M(z_1,z_0).
\label{eq:C5}
\end{equation}
The X are operators in the Lie algebra of SU(2)
and are independent of \A{Z}{z} .  $z_{n+1}\ and\ z_0 $ are the
boundary points, and for simplicity I suppress the integrations
over the remaining $dz_i $.  Multiply $\psi $ by the following
partition of unity.
\begin{equation}
1 = [\Theta (z_{n+1}, z, z_n) + \Theta (z_n, z, z_{n-1}) + \cdots
+
      \Theta (z_2, z, z_1) + \Theta (z_1, z, z_0) ].
\label{eq:C6}
\end{equation}
Then
\begin{eqnarray}
     [1 /\E{z}{Z}] \psi &=& [1 /\E{z}{Z}] [\Theta (z_{n+1}, z, z_n)
     + \cdots + \Theta (z_2, z, z_1) + \Theta (z_1, z, z_0) ] \psi
               \nonumber \\
     &=& i(\hbar S_Z)^{-1} \Theta (z_{n+1}, z ,z_n) M(z_{n+1}, z_n)
               X(z_n) \cdots M(z_1,z_0) \nonumber \\
     & +& \cdots + M(z_{n+1}, z_n) X(z_n) \cdots
          i(\hbar S_Z)^{-1} \Theta (z_1, z, z_0) M(z_1,z_0),
\label{eq:C7}
\end{eqnarray}
where \EQ{C4} was used on the second line.  Provided no $S_Z $
vanishes, this action is completely finite, QED.

\section{The Area Operator}

     The classical expression for the area of a surface S with
normal $n_a$ is
\begin {equation}
     A = \int_S d^2\sigma \sqrt{\E{a}{I} \E{b}{I} n_a n_b}.
\label{D1}
\end{equation}
In order to regulate this product of operators, Rovelli and
Smolin \cite{AV} evaluate the \Etld\ operators at two slightly
separated
points $\alpha (s)\ and\ \alpha (t) $ on a small loop $\alpha $,
then insert holonomies between the separated \Etld\ operators to
maintain gauge invariance.  In
order to obtain diffeomorphism invariance, they square A and
doubly-
integrate over the surface.  The result is
\begin{equation}
     A^2 = \int_S d^2\sigma \int_S d^2\tau |(1/2)c^2
                    T^{ab}n_a n_b|,
\label{eq:D2}
\end{equation}
where
\begin{equation}
     T^{ab}[\alpha](s,t) = Tr[M(s,t) \E{a}{I}(\alpha (t))
(\sigma^I /ic) M(t,s) \E{b}{J}(\alpha (s)) (\sigma^J /ic)],
\label{eq:D3}
\end{equation}
and the holonomies M are integrated along the small loop
$\alpha $.  Rovelli and Smolin renormalize all their Pauli
$\sigma^I $ matrices with (an arbitrary real factor c times) a
factor i, as in \EQ{D3}.  This is done primarily to remove an
explicit factor of i from the holonomy; however,  Rovelli and
Smolin then have to contend with two factors of  i in the
definition of the area operator.  They have removed  these from
\EQ{D2} by
using absolute value bars rather than a factor (-1).  As a
result, the $\alpha \rta 0 $ limit of \EQ{D3} contains $|\E{a}{I}
\E{b}{I} |$ rather than simply  \E{a}{I} \E{b}{I}.  The absolute
value
bars are signifigant and I shall return to them later.

     First, however, one must ask how the above, 3+1
dimensional construction must be modified to fit the
 planar case.  It is possible to define area operators
for areas lying in the xz, yz, and xy planes.  Since the $\psi $
are functionals of only a single connection \A{Z}{z}, however,
only the xy area operator $\propto \E{z}{Z} $ is a simple
functional derivative.  The other two area operators multiply
$\psi $ by a function, and there is no hope of a simple action.
I will therefore concentrate on the xy area operator.  The
construction of this operator is immediate: replace the general
holonomies in \EQ{D3} by holonomies along the z-axis; replace
both \E{a}{I} by \E{z}{Z}; replace the factor ic by my
normalization, a
factor of 2.

     Since the factors of i have now disappeared from $T^{ab}$,
there is no need for absolute
value bars (or a factor (-1)) in \EQ{D2}.  The resultant $A^2$
operator has a small $\alpha $ limit $\propto \E{z}{Z}^2 $, the
limit  Rovelli and Smolin would have obtained had they used a
factor (-1) rather than absolute value bars.

     Now for simplicity, and to avoid the complexities of the
Penrose calculus,  let $A^2$ act upon the holonomy constructed
from the j = 1/2 representation.  This holonomy is infinitely
functionally differentiable, so that the point-splitting in
\EQ{D3} is overkill, and one may replace the holonomies in
$T^{ab}$ by unit matrices.  Then the area operator collapses to
\begin{equation}
     A^2 = [\int dxdy \E{z}{Z}]^2.
\label{eq:D4}
\end{equation}
Each \E{z}{Z} brings down a factor of $-i \hbar S^Z $ (or  $\hbar
\tau^Z $ if one uses the Rovelli-Smolin normalization, with the
factor of i hidden in $\tau^Z \propto \sigma^Z /i)$.  Thus $A^2 $
has a negative eigenvalue:
\bea
     A^2 M &=& (-i \hbar S^Z )^2 M \nonumber \\
          &=& -(\hbar S^Z )^2 M.
\label{eq:D5}
\eea
The eigenvalue $(S^Z)^2 $ is 1/4, only 1/3 of the Rovelli-Smolin
value j(j+1) = 3/4, but the 1/3 can be accounted for by the shift
from three spatial dimensions to one.  The factor of (-1) is the
important feature: it comes from the factor of i in the holonomy,
and has nothing to do with dimension.  This factor would be
removed by the absolute value bars, but it is not clear that the
absolute value bars are demanded by any fundamental principle of
the theory.

     This difficulty can be made more striking.  Note that in the
planar
case  it is possible to take the square root of the $A^2 $
operator and still get a diffeomorphism invariant result.  The
square root
\begin{equation}
     A = \int dx dy \E{z}{Z}
\label{eq:D6}
\end{equation}
is also gauge invariant, since the only surviving gauge rotation
(around the Z axis) leaves the Z index invariant.  This operator
has an {\it imaginary} eigenvalue when acting on the
holonomy.

\end{document}